\documentclass[12pt,a4paper]{article}

\def\Ra{\Rightarrow}

\def\({\left(}
\def\){\right)}
\def\[{\left[}
\def\]{\right]}

\def\a{\alpha}
\def\be{\beta}

\def\de{\delta}

\def\la{\lambda}

\def\om{\omega}
\def\Om{\Omega}

\def\coma{\ ,\quad}
\def\build#1_#2^#3{\mathrel{\mathop{\kern 0pt#1}\limits_{#2}^{#3}}}

\def\beq{\begin{equation}}
\def\eeq{\end{equation}}
\def\beqa*{\begin{eqnarray*}}
\def\eeqa*{\end{eqnarray*}}
\def\ba{\hspace{-7mm}\begin{array}{lll}}
\def\ea{\end{array}}
\newcommand\bm[1]{\mbox{\boldmath$#1$}}

\usepackage{amsfonts,amsmath,amssymb}
\usepackage{mathrsfs}
\usepackage{amscd}
\usepackage{color}
\usepackage{graphicx}

\begin{document}
\title{Perturbation theory and harmonic gauge propagation in general relativity, a particular example}
\author{Jes\'us Mart\'{\i}n\thanks{Instituto Universitario de F\'isica Fundamental y Matem\'aticas (IUFFyM), Universidad de Salamanca, Spain, e-mail address: chmm@usal.es},  Alfred Molina\thanks{Dept. F\'{\i}sica Fonamental, Institut de Ci\`encies del Cosmos, Universitat de Barcelona, Spain, e-mail address: alfred.molina@ub.edu} and Eduardo Ruiz\thanks{Depto. de F\'isica Fundamental, Facultad de Ciencias, Universidad de Salamanca, Spain, e-mail address: eruiz@usal.es}}
\date{}
\maketitle

\begin{abstract}
We study  how the changes of coordinates between the class of harmonic coordinates affect the analitycal solutions of Einstein's equations and we apply it to an analytical approach for stationary and axisymmetric solutions of Einstein equation used
by the authors \cite{cmmr}, \cite{cugmmr} to solve the problem of a self-gravitating rigidly rotating perfect fluid compact source.
\end{abstract}


\section{Introduction\label{sec1}}
In several previous articles \cite{cmmr,cugmmr} to solve Einstein's equations we used a new  method to obtain approximate 
stationary and axisymmetric solutions to describe the gravitational field (interior
and exterior) of a compact stellar object in rigid rotation. This method is a combination of a postminkowskian approach and another one that could be qualified as post spherical. At every postminkowskian order in a dimensionless parameter, $\lambda$ that we  set as usual
to be proportional to the mass of the object, a series expansion is performed  in powers of a parameter, $\Omega$, also dimensionless,
which takes into account of the  deformation of the object relative to the spherical symmetry. This parameter classically represent the ratio in the equatorial plane between the centrifugal force and the gravitational one, or in terms of the Maclaurin ellipsoids \cite{Chandrasekhar1,Chandrasekhar2}, some function of the eccentricity. In this sense this second subordinate series expansion is similar to the slow rotation approach used by Hartle \cite{Hartle}, but with the advantage that we do not start from an spherical exact solution in $\lambda$ and then we can go easily to higher $\Omega$ terms and we do not need any numerical computation to obtain algebraic results in the slow rotation parameter.

In this article we are going to study only some aspects related to the postminkowskian approach without any care with respect to the post spherical approach.
  
To explain our current goal let's recall first that the approximate general solution of the Einstein equations to the postminkowskian linear order in harmonic coordinates contains four groups of constants. As regards the exterior solution two of these groups of constants are the static and dynamic multipole moments of Geroch-Hansen, Thorne et al \cite{Thorne1,Geroch, Hansen} (and therefore they are invariant), while the other two groups are removable by a coordinate change, so that they can be described by {\it{gauge constants}\/}. On the interior a similar situation occurs, there are two sets of non removable constants, but they do not have a specific meaning as for the exterior solution, and two other groups gauge constants. A priori it seems natural to choose a coordinate system in which all gauge constants become equal to zero, this is the choice of Thorne\cite{Thorne2} and that is known as {\it canonical gauge\/}. However we are interested in making a match on the surface of the stellar object by using the Lichnerowicz's prescription\cite{Lichnerowicz}, i.e. in this matching on the surface the metric is continuous and has continuous derivatives. On this basis the gauge constants (both external and internal) are needed and also are uniquely fixed when the matching is performed\cite{cugmmr}.

As a consequence of this we need to maintain the gauge constants at the first order and drag them to the second order (and so on) in an iterative process that suppose the postminkowskian approach. It is easy to see that this result in quite long and cumbersome calculations, even with the use of the algebraic software, which must be implemented with elaborate routines when we go over the second order. The following questions can be naturally raised:

\begin{itemize}
\item What is the aspect of the solution at the second order (for instance) that comes from the first order gauge  constants?

\item You may obtain this part of the second order solution just analyzing infinitesimal coordinate changes up to the second order?
\end{itemize}

The main objective of this paper is to answer these questions, showing that the global solution to the second order contains the first order gauge constants transferred to the second order solution that matches exactly with we expect from the analysis of infinitesimal coordinate changes. One might object that the result is trivial because of the covariance of the theory; however this argument present some difficulties which are specified in the text, and in any case our contribution is a explicit verification of how gauge constants are transferred, which can simplify the calculations significantly.

The paper is organized as follows. In Section 2 the expressions for the harmonicity conditions and the Ricci tensor are written in an appropriate way to develop the postminkowskian approximation. In Section 3, these expressions are used to write Einstein's equations to different orders in the postminkowskian development. In Section 4 the infinitesimal coordinate changes are studied up to the second order including the peculiarities of the {\it harmonic changes \/}, i.e., those who go from harmonic coordinates to harmonic coordinates, specially the results are written for our problem i.e. for an spacetime stationary and axisymmetric with a Papapetrou structure and asymptotically flat.
In Section 5  the properties of the stationary and axisymmetric metrics are specified. In Section 6 we address the problem of external vacuum solution, solving Einstein's equations to first and second order of the postminkowskian approximation in harmonic coordinates. Besides we prove the fundamental thesis of this work, i.e., the part of the second order solution transferred by the first order gauge constants  can be obtained by a simple analysis of the infinitesimal changes. It is also argued that the result must be true when we are going to the third order. In Section 7 the same process is repeated for the case of an interior  perfect fluid solution in a rigid rotation and with a linear equation of state. Finally in Section 8 the conclusions of the work are presented.

As usual the space--time metric is written as follow
$$
ds^2 = g_{\a\be}(x^\rho) dx^\a dx^\be
$$
The Greek and Latin index take the values
$$
\a,\be,\la,\ldots = 0,1,2,3\ ;\qquad i,j,k,\ldots =1,2,3
$$
We use the Einstein summation convention and the signature of the space--time is $\big(-,+,+,+\big)$.


\section{Harmonicity conditions and Ricci tensor \label{sec2}}
\subsection{Harmonicity conditions}
It is well known that the harmonic coordinates condition is
\begin{equation}
\Gamma^\alpha := g^{\lambda\mu}\Gamma^\alpha_{\lambda\mu} = -\frac1{\sqrt{-g}}
\partial_\mu\(\sqrt{-g}g^{\mu\alpha}\) = 0
\end{equation}
but
\begin{equation}
\Gamma^\rho_{\mu\rho} = \partial_\mu\log\sqrt{-g}
\end{equation}
and then we have
\begin{equation}
\Gamma^\a = -\partial_\mu g^{\mu\alpha} - g^{\mu\alpha} \Gamma^\rho_{\mu\rho}\label{eq1} 
\end{equation}
We can define also
\begin{equation}
\Gamma_\alpha:= g_{\alpha\lambda}\Gamma^\lambda = -g_{\alpha\lambda}\partial_\mu g^{\mu\lambda} -
\Gamma^\mu_{\alpha\mu}\label{eq2}
\end{equation}

If we define the metric deviation $h_{\alpha\beta}$ from Minkowski metric as
\begin{equation}
g_{\alpha\beta} := \eta_{\alpha\beta} + h_{\alpha\beta}\label{metdev}
\end{equation}
and the inverse metric
\begin{equation}
g^{\alpha\beta}:= \eta^{\alpha\beta} + k^{\alpha\beta}\quad\big(\Ra\ k^{\alpha\rho}\eta_{\rho\beta} + \eta^{\alpha\rho}h_{\rho\beta}
+k^{\alpha\rho}h_{\rho\beta} =0\big)\label{metdev1}
\end{equation}
the equation (\ref{eq2}) can be written in terms of the metric deviation as follows
\begin{equation}
\Gamma_\alpha=\eta^{\lambda\mu}\Gamma_{\alpha,\lambda\mu} + k^{\lambda\mu}\Gamma_{\alpha,\lambda\mu}= l_\alpha + P_\alpha =0
\label{eq4}
\end{equation}
where  
\begin{equation}
\begin{aligned}
&l_\alpha:= \partial^\mu h_{\mu\alpha} - \frac12 \partial_\alpha h
\\[1ex]
&P_\alpha :=  k^{\lambda\mu}\Gamma_{\alpha,\lambda\mu}
\end{aligned}
\qquad\big(\partial^\mu:=\eta^{\mu\rho}\partial_\rho\ ,\quad  h:= \eta^{\lambda\mu}h_{\lambda\mu}\big)
\end{equation}
by splitting the linear terms in the deviation from the nonlinear terms.

\subsection{Ricci tensor}
Now we are going to  write the Ricci tensor in terms of the left hand side of harmonicity condition and the metric deviation. First of all from the definition of the Ricci tensor we have
\begin{equation}
R_{\alpha\beta} = \partial_\lambda\Gamma^\lambda_{\alpha\beta}-
\partial_\beta\Gamma^\lambda_{\alpha\lambda}+ \Gamma^\lambda_{\rho\lambda}
\Gamma^\rho_{\alpha\beta} -\Gamma^\lambda_{\rho\beta} \Gamma^\rho_{\alpha\lambda}\label{Ricci}
\end{equation}
but using \eqref{eq1} the first term of the right hand  can be written
\begin{align}
\partial_\lambda\Gamma^\lambda_{\alpha\beta} &=  g^{\lambda\mu}\partial_\lambda\Gamma_{\mu,\alpha\beta} + \partial_\lambda
g^{\lambda\mu}\Gamma_{\mu,\alpha\beta}\notag
\\[1ex]
&= g^{\lambda\mu}\partial_\lambda\Gamma_{\mu,\alpha\beta} -\Gamma^\mu
\Gamma_{\mu,\alpha\beta} - \Gamma^\lambda_{\rho\lambda}\Gamma^\rho_{\alpha\beta}
\end{align}
and using this result in \eqref{Ricci}
\begin{equation}
R_{\alpha\beta} = g^{\lambda\mu}\partial_\lambda\Gamma_{\mu,\alpha\beta} -\partial_\beta\Gamma^\lambda_{\alpha\lambda} -
\Gamma_\mu\Gamma^\mu_{\alpha\beta} -\Gamma^\lambda_{\rho\beta} \Gamma^\rho_{\alpha\lambda}
\end{equation}
In this expression only the first and the second term in the right hand side contain linear terms in the deviation. We are going to split them from the non linear ones; since  
\begin{align}
g^{\lambda\mu}\partial_\lambda\Gamma_{\mu,\alpha\beta} - \partial_\beta\Gamma^\lambda_{\alpha\lambda} &=
\frac12\eta^{\lambda\mu}\(\partial_{\lambda\alpha}h_{\mu\beta} + \partial_{\lambda\beta}h_{\mu\alpha}
- \partial_{\lambda\mu}h_{\alpha\beta}\)\notag
\\[1ex]
& + k^{\lambda\mu}\partial_\lambda\Gamma_{\mu,\alpha\beta}- \frac12\eta^{\lambda\mu} 
\partial_{\beta\alpha}h_{\lambda\mu} -\partial_\beta(k^{\lambda\mu}\Gamma_{\mu,\alpha\lambda}) \
\end{align}
then the Ricci tensor can be expressed as
\begin{align}
R_{\alpha\beta} =  L_{\alpha\beta} - \Gamma_\mu\Gamma^\mu_{\alpha\beta} + N_{\alpha\beta} 
\end{align}
where we have defined
\begin{align}
&L_{\alpha\beta} := -\frac12\square h_{\alpha\beta}
+\frac12\partial_\alpha l_\beta +
\frac12\partial_\beta l_\alpha
\\[1.5ex]
&N_{\alpha\beta} := k^{\lambda\mu}\partial_\lambda\Gamma_{\mu,\alpha\beta}
-\frac12\partial_\beta(k^{\lambda\mu}\partial_\alpha h_{\lambda\mu}) -\Gamma^\lambda_{\rho\beta} \Gamma^\rho_{\alpha\lambda}
\end{align}
and where all the linear terms in the deviation are in $L_{\alpha\beta}$. If we use the harmonicity condition $\Gamma_\alpha=0$ we can substitute  $l_\alpha= -P_\alpha$
and we finally get
\begin{equation}
R_{\alpha\beta} = -\frac12\square h_{\alpha\beta} -\frac12\partial_\alpha P_\beta -
\frac12\partial_\beta P_\alpha + N_{\alpha\beta} \label{eq5}
\end{equation}
where only the first term in the right hand side contains linear terms in the deviation.


\section{Einstein equations and series development\label{sec3}}
The Einstein equations read
\begin{equation}
R_{\alpha\beta} = T_{\alpha\beta}-\frac12 T g_{\alpha\beta} := {\cal T}_{\alpha\beta}\quad(8\pi G = c = 1)
\end{equation}
Let us assume that the deviation $h_{\a\be}$ of the metric and ${\cal T}_{\alpha\beta}$ can be developed in a series of a parameter $\la$, i.e.
\beq
\left\{\begin{aligned}
&h_{\a\be} = \la h^{(1)}_{\a\be} +  \la^2 h^{(2)}_{\a\be} 
+  \la^3 h^{(3)}_{\a\be} +\cdots \,= \sum_{n=1}\la^n h^{(n)}_{\alpha\beta}
\\[1ex]
&{\cal T}_{\alpha\beta}= \la{\cal T}^{(1)}_{\a\be} +  \la^2 {\cal T}^{(2)}_{\a\be} 
+  \la^3 {\cal T}^{(3)}_{\a\be} +\cdots \,= \sum_{n=1}\la^n {\cal T}^{(n)}_{\alpha\beta}
\end{aligned}\right.
\eeq
Then the linear term as in the harmonic coordinates condition (\ref{eq4}) can be written 
\begin{equation}
l_\alpha=\sum_{n=1}\la^n\big[ \partial^\beta h^{(n)}_{\alpha\beta}-\frac12 \partial_\alpha h^{(n)}\big]:=\sum_{n=1}\la^n l^{(n)}_\alpha\, ,\quad h^{(n)}:=\eta^{\mu\nu} h^{(n)}_{\mu\nu}
\end{equation}
For the nonlinear term  we have
$$
P_\alpha =  \la^2 P^{(2)}_\a +  \la^3 P^{(3)}_\a 
+\cdots  =  \sum_{n=2} \la^n P^{(n)}_\a := \sum_{n=2} \la^n\!\sum_{r+s=n} \!k^{(r)\mu\nu} \,\Gamma^{(s)}_{\alpha,\mu\nu}
$$
$(r,s=1,2,\dots)$ where in  $\Gamma^{(s)}_{\alpha,\lambda\mu}$ only the terms $\partial_\nu h^{(s)}_{\beta\rho}$ are needed. The terms $k^{(r)\mu\nu}$ of  the inverse metric are obtained as follows; from the definition
$$
k^{\alpha\beta} = \sum_{n=1}\la^n k^{(n)\alpha\beta}
\quad  \mbox{and the relation}\quad g^{\alpha\mu}g_{\mu\beta}=\delta^\alpha_\beta
$$
we obtain 
$$\sum_{n=1}\la^n\left[k^{(n)\alpha\rho}\eta_{\rho\beta} + \eta^{\alpha\rho}h^{(n)}_{\rho\beta}
+\sum_{r+s=n}k^{(r)\alpha\rho}h^{(s)}_{\rho\beta}\right] =0
$$
If we know $h^{(1)}_{\mu\nu}$   this gives us $k^{(1)\mu\nu}$  and if we know
$$
\{h^{(1)}_{\mu\nu},\dots,h^{(n)}_{\mu\nu}\}\quad  \mbox{and}\quad \{k^{(1)\mu\nu},\dots,k^{(n-1)\mu\nu}\}
$$
the term with $\la^n$ give us the term $k^{(n)\mu\nu}$. 
 Then to obtain the term $P^{(n)}_\alpha$
we need the deviation of the metric to order $n-1$.

The same happen with the Ricci tensor (\ref{eq5}). The remaining terms to analyze are in $N_{\alpha\beta}$, in which the two first terms have the same structure we have already analyzed, and the third one contains products of the type
$$(\eta^{\lambda\mu}+k^{\lambda\mu})\Gamma_{\mu,\alpha\rho}(\eta^{\rho\nu}+k^{\rho\nu})\Gamma_{\nu,\beta\lambda}
$$
The terms of order $n$ are obtained with the deviation of the metric to order less than order $n$, and like $P_\alpha$ in $N_{\alpha\beta}$ the lower term  is of order $\la^2$, i.e.
\begin{equation}
N_{\alpha\beta} =  \la^2 N^{(2)}_{\a\be} + \la^3 N^{(3)}_{\a\be} 
+\cdots= \sum_{n=2}\la^i N^{(n)}_{\alpha\beta} 
\end{equation}

Then to first order the Einstein's equation and the harmonicity conditions are
\beq
\left\{\ba
&&\square\, h^{(1)}_{\a\be} = -2\, {\cal T}^{(1)}_{\a\be}
\\[2ex]
&&\partial^\rho h^{(1)}_{\rho\a} -\dfrac12\, \partial^\a h^{(1)} = 0
\ea\right.
\eeq
and to order  $\bm{n\ge 2}$
\beq
\left\{\ba
&&\square\, h^{(n)}_{\a\be} = -2\, {\cal T}^{(n)}_{\a\be} +2\left[N^{(n)}_{\a\be} - \partial_{(\a}P^{(n)}_{\be)}\right]
 \\[3ex]
&&\partial^\rho h^{(n)}_{\rho\a} -\dfrac12\,\partial^\a h^{(n)} = -P^{(n)}_\a
\ea\right.
\eeq
where for instance for $n=2$
${\cal T}^{(2)}_{\alpha\beta}$, $N^{(2)}_{\alpha\beta}$ and $P^{(2)}_\alpha$ are build with the linear deviation  $h^{(1)}_{\alpha\beta}$ i.e.  
\begin{equation}
\begin{aligned}
& P^{(2)}_\alpha = k^{(1)\lambda\mu} \Gamma^{(1)}_{\alpha,\lambda\mu}
\\[1ex]
& N^{(2)}_{\alpha\beta} = k^{(1)\lambda\mu} \partial_\lambda\Gamma^{(1)}_{\mu,\alpha\beta} - \frac12 k^{(1)\lambda\mu}\partial_{\alpha\beta} h{(1)}_{\lambda\mu} -\frac12 \partial_\alpha h^{(1)}_{\lambda\mu}\partial_\beta k^{(1)\lambda\mu}-{\Gamma}^{(1)\rho}_{\beta\lambda}\label{eqPN} {\Gamma}^{(1)\lambda}_{\alpha\rho}
\end{aligned}
\end{equation}
and ${\cal T}^{(2)}_{\alpha\beta}$ is build from $h^{(1)}_{\lambda\mu}$. The same happen with the higher order equations,  the second member of the  equations are build with lower orders that are already known.
\section{Infinitesimal coordinates change to second order\label{sec4}}
Now we are going to study how the infinitesimal changes of coordinates affect the deviation of the metric of the previous section and how this gauge propagates to higher order.
Let us assume that we take the deviation to second order in the parameter, i.e.  
\begin{equation}
g_{\alpha\beta}(x) = \eta_{\alpha\beta} + \la h_{\alpha\beta}(x) + \la^2 q_{\alpha\beta}(x)
\end{equation}
and the inverse metric
\begin{equation}
g^{\alpha\beta}(x) = \eta^{\alpha\beta} + \la k^{\alpha\beta}(x) + \la^2 p^{\alpha\beta}(x)
\end{equation}
Let us point out that we have changed the notation, the terms  $h_{\alpha\beta}$ and $k^{\alpha\beta}$ represent from now on the first order in the parameter $ \la$ for  the deviation and not all the deviation as in equations  (\ref{metdev}) and (\ref{metdev1}). i.e.
$$
\left\{\begin{aligned}
&h^{(1)}_{\alpha\beta}\rightarrow h_{\alpha\beta}\coma h^{(2)}_{\alpha\beta}\rightarrow q_{\alpha\beta}
\\[1ex]
&k^{(1)\alpha\beta}\rightarrow k^{\alpha\beta}\coma k^{(2)\alpha\beta}\rightarrow p^{\alpha\beta}
\end{aligned}\right.
$$
Then from (\ref{metdev1})
\begin{equation}
k^{\alpha\rho}\eta_{\rho\beta} + \eta^{\alpha\rho}h_{\rho\beta}=0\ ; \quad
p^{\alpha\rho}\eta_{\rho\beta} + \eta^{\alpha\rho}q_{\rho\beta}+k^{\alpha\rho}h_{\rho\beta} =0\label{eqmetinv}
\end{equation}
If we perform now the infinitesimal coordinates change to second order
\begin{equation}
\tilde x^\alpha = x^\alpha +\la\,\xi^\alpha(x) + \la^2\zeta^\alpha(x)
\end{equation}
the metric in the new coordinates can be written as
\begin{align}
\tilde g_{\alpha\beta}(\tilde x) &= \eta_{\alpha\beta} + \la \tilde h_{\alpha\beta}(\tilde x) + \la^2 \tilde q_{\alpha\beta}(\tilde x)
\\[1ex]
&=\eta_{\alpha\beta} + \la\tilde h_{\alpha\beta}(x)+ \la^2\Big[\xi^\rho(x)\partial_\rho\tilde h_{\alpha\beta}(x) +  \tilde q_{\alpha\beta}(x)\Big]\label{newmet}
\end{align}
but also
\begin{equation}
g_{\alpha\beta}(x) = \frac{\partial\tilde x^\lambda}{\partial x^\alpha}(x) \frac{\partial\tilde x^\mu}{\partial x^\beta}(x) \tilde g_{\lambda\mu}\big[\tilde x(x)\big]
\end{equation}
 and then
\begin{eqnarray}
\eta_{\alpha\beta} + \la h_{\alpha\beta}(x)\!\!\!\! &+&\!\!\!\! \la^2 q_{\alpha\beta}(x)= (\delta^\lambda_\a+\la\partial_\alpha\xi^\lambda + \la^2\partial_\alpha\zeta^\lambda)
(\delta^\mu_\beta+\la\partial_\beta\xi^\mu+ \la^2\partial_\beta\zeta^\mu)\nonumber
\\[1.5ex]
&\times&\!\!\!\! \left\{\eta_{\lambda\mu} + \la\tilde h_{\lambda\mu}(x)+ \la^2\Big[\xi^\rho(x)\partial_\rho\tilde h_{\lambda\mu}(x) +  \tilde q_{\lambda\mu}(x)\Big]\right\}\label{chcor}
\end{eqnarray}
Now we can identify the two first orders in $\la$ in equations (\ref{newmet}) and (\ref{chcor})
\begin{equation}
\tilde h_{\alpha\beta}(x) = h_{\alpha\beta}(x)-\partial_\alpha\xi_\beta(x) -\partial_\beta\xi_\a(x)\label{eq6}
\end{equation}
\begin{align}
\tilde q_{\alpha\beta}(x) &= q_{\alpha\beta}(x) - \partial_\alpha\zeta_\beta(x) -\partial_\beta\zeta_\a(x) - \partial_\alpha\xi^\mu(x) \partial_\beta\xi_\mu(x)\notag
\\[1ex]
&- \xi^\rho(x)\partial_\rho\Big[ h_{\alpha\beta}(x)-\partial_\alpha\xi_\beta(x) -\partial_\beta\xi_\a(x)\Big]\notag
\\[1ex]
&
-\Big[ h_{\alpha\mu}(x)-\partial_\alpha\xi_\mu(x) -\partial_\mu\xi_\a(x)\Big]
\partial_\beta\xi^\mu(x)\notag
\\[1ex]
& - \Big[ h_{\mu\beta}(x)-\partial_\beta\xi_\mu(x) -\partial_\mu\xi_\beta(x)\Big]\partial_\alpha\xi^\mu(x)
\end{align}
which can be written as
\begin{equation}
\tilde q_{\alpha\beta}(x) = q_{\alpha\beta}(x) - \partial_\alpha\zeta_\beta(x) -\partial_\beta\zeta_\a(x) -\pounds(\mbox{\boldmath{$\xi$}})\tilde h_{\alpha\beta}(x) - \partial_\alpha\xi^\mu(x) \partial_\beta\xi_\mu(x)\label{eq7}
\end{equation}
where the two last terms are the first order gauge propagation to second order.

To use it later on it is important to specify the equations that should verify the  vectors $\xi^\a(x)$ and $\zeta^\a(x)$  
to accomplish that the change (31) transforms harmonic coordinates to harmonic coordinates (``harmonics changes"). As is known it is necessary that
\beq
g^{\la\mu}\partial_{\la\mu}\tilde x^\a(x) =0
\eeq
which results in the following equations
\beq
\begin{aligned}
&\eta^{\la\mu}\partial_{\la\mu}\xi^\a(x) =0
\\[1.5ex]
&\eta^{\la\mu}\partial_{\la\mu}\zeta^\a(x) =  h^{\la\mu}(x)\partial_{\la\mu}\xi^\a(x)\label{eq40}
\end{aligned}
\eeq
having taken into account (29), (30) and (31), and where $h^{\la\mu}=\eta^{\la\a}\eta^{\mu\be} h_{\a\be}\,$.

\section{Stationary and axisymmetric metrics\label{sec5}}
We require space-time to be a stationary and axisymmetric Riemannian asymptotically flat manifold admitting a global system of  spherical-like coordinates $\{t,r,\theta,\varphi\}$ which verifies the following properties:

\noindent {\bf A)} Coordinates are adapted to the space-time symmetry, that is to say, $\bm\xi=\partial_t$ and $\bm\zeta=\partial_\varphi$ are respectively the timelike and spacelike Killing vectors, so that the metric components do not depend on coordinates $t$ or $\varphi$.

\noindent {\bf B)}  Coordinates $\{r,\theta\}$ parametrize two dimensional surfaces orthogonal to the orbits of the symmetry group, that is the metric
tensor has Papapetrou structure,
\begin{eqnarray}
&\bm{g} = \gamma_{tt}\bm{\omega}^t{\otimes}\bm{\omega}^t
+\gamma_{t\varphi}(\bm{\omega}^t{\otimes}\bm{\omega}^\varphi+\bm{\omega}^\varphi{\otimes}\bm{\omega}^t)+
\gamma_{\varphi\varphi}\bm{\omega}^\varphi{\otimes}\bm{\omega}^\varphi
\nonumber\\
&\quad +\gamma_{rr}\bm{\omega}^r{\otimes}\bm{\omega}^r+
\gamma_{r\theta}(\bm{\omega}^r{\otimes}\bm{\omega}^\theta+\bm{\omega}^\theta{\otimes}\bm{\omega}^r) 
+\gamma_{\theta\theta}\bm{\omega}^\theta{\otimes}\bm{\omega}^\theta,
\label{eqmetrica}
\end{eqnarray}
where 
$\bm{\omega}^t=dt$, $\bm{\omega}^r=dr$, $\bm{\omega}^\theta=r\,d\theta$, $\bm{\omega}^\varphi=r\sin\theta\,d\varphi$ 
is the Euclidean orthonormal co basis associated to these coordinates.

\noindent {\bf C)} Coordinates $\{t,x=r\sin\theta\cos\varphi,\,y=r\sin\theta\sin\varphi,\,z=\cos\theta\}$ associated with the spherical-like coordinates are Cartesian coordinates at spacelike infinity, that is the metric in these coordinates tends to the Minkowsky metric in standard Cartesian coordinates for large values of the coordinate $r$.

All these properties are compatible with the harmonic coordinates we use in this paper. Time coordinate $t$ is always harmonic under these assumptions.

The coordinate changes  to another system of adapted coordinates $\{t',\varphi'\}$ preserving the regularity in the axis of symmetry and with closed compact orbits of periodicity $2 \pi$ for the axial Killing vector are $t=a\, t'$ and $\varphi=\varphi'+b\,t'\,$, where $a$ and $b$ are constants. If at spacelike infinity the metric tends to the Minkowski metric then $a=1$ and this implies for the infinitesimal change of Section \ref{sec4} that $\xi^0(x)=0,\zeta^0(x)=0$, and taking also into account the independence of the metric on time coordinate we have $\xi^i(x^j)$ and $\zeta^i(x^k)$, i.e. do not depend on time coordinate. Furthermore the change $\varphi=\varphi'+bt'$ do not maintain the harmonicity condition, i.e. if the Cartesian coordinates $x,y,z$ associated to the spherical $r,\theta,\varphi$ are harmonic the Cartesian ones associated to $r,\theta,\varphi'$ are not harmonic.

Consequently, splitting time and space components in the equations (\ref{eq6}) and (\ref{eq7}) we have\\[1ex]
{\bf First order}
\begin{align}
&\tilde h_{00}(x^k) = h_{00}(x^k)\label{eqinf1}
\\[1ex]
&\tilde h_{0j}(x^k) = h_{0j}(x^k)\label{eqinf2}
\\[1ex]
&\tilde h_{ij}(x^k) = h_{ij}(x^k)-\partial_i\xi_j(x^k)- \partial_j\xi_i(x^k)\label{eqinf3}
\end{align}
the metric components $h_{00}$ y $h_{0j}$ of the metric deviation are invariants to first order.

\noindent{\bf Second order}
\begin{align}
&\tilde q_{00}(x^k) = q_{00}(x^k) - \xi^l(x^k)\partial_l h_{00}(x^k)\label{eqinf4}
\\[1ex]
&\tilde q_{0j}(x^k) = q_{0j}(x^k) 
- \xi^l(x^k)\partial_l h_{0j}(x^k) - h_{0l} \partial_j\xi^l(x^k)\label{eqinf5}
\\[1ex]
&\tilde q_{ij}(x^k) = q_{ij}(x^k)-\partial_i\zeta_j(x^k) - \partial_j\zeta_i(x^k)-\pounds(\mbox{\boldmath{$\xi$}})\tilde h_{ij}(x^k) - \partial_i\xi^l(x^k) \partial_j\xi_l(x^k)\label{eqinf6}
\end{align}

With respect to the ``harmonics changes" (\ref{eq40}), the stationarity and axial symmetry with a Papapetrou lead to the following equations
\begin{align}
&\Delta\xi^i(x) =0
\\[1.5ex]
&\Delta\zeta^i(x) =  h^{jk}(x)\partial_{jk}\xi^i(x)\label{eq49}
\end{align}
i.e. the functions $\xi^i$  must be three independent harmonic functions. On the other hand, the functions $\zeta^i$ must be the sum of a harmonic function and a particular solution of equation \eqref{eq49}.

Finally, the harmonic condition and Einstein's equations of Section \ref{sec3} with the splitting of time and space components become:

\noindent{\bf First order}
\begin{align}
&\Delta h^{(1)}_{00} = -2 {\cal T}^{(1)}_{00}\label{eq8}
\\[1ex]
&\Delta h^{(1)}_{0j} = -2 {\cal T}^{(1)}_{0j}\,;\quad
\partial^j h^{(1)}_{0j}  = 0\label{eq9}
\\[1ex]
&\Delta h^{(1)}_{ij} = -2 {\cal T}^{(1)}_{ij}\,;\quad
\partial^j h^{(1)}_{ij}-\frac12\partial_j \hat{h}^{(1)}=-\frac12 \partial_j h^{(1)}_{00}  \label{eq10}
\end{align}
where $\hat{h}:=\delta^{ij} h_{ij}$.

\noindent{\bf Order $\bm{n\ge 2}$}
\begin{align}
&\Delta h^{(n)}_{00} = -2 {\cal T}^{(n)}_{00} +2N^{(n)}_{00}\label{eq11}
 \\[1ex]
&\Delta h^{(n)}_{0j} = -2 {\cal T}^{(n)}_{0j} +2N^{(n)}_{0j} - \partial_jP^{(n)}_0\,
 ;\quad \partial^j h^{(n)}_{0j} = -P^{(n)}_0\label{eq12}
\\[1ex]
&\Delta h^{(n)}_{ij} = -2 {\cal T}^{(n)}_{ij} +2\left[N^{(n)}_{ij} - \partial_{(i}P^{(n)}_{j)}\right]
 \,;\quad \partial^j h^{(n)}_{ij} -\dfrac12\partial_i \hat{h}^{(n)} = -P^{(n)}_i-\frac12 \partial_i h^{(n)}_{00}\label{eq13}
\end{align}


\section{Vacuum exterior solution in harmonic\\ coordinates}

Let's assume that we have a compact stationary gravitational source with axial symmetry and we would like to study the exterior 2--postminkowskian solution \cite{cmmr}. First of all we need to solve equations (\ref{eq8}), (\ref{eq9}) and (\ref{eq10}) with ${\cal T}^{(1)}_{\mu\nu}=0$, and then to solve the equations (53), (54) and (55) for $n=2$ and ${\cal T}^{(2)}_{\mu\nu}=0$. Henceforth we will use the notations of (28) and (29).

\subsection{First order in $\la$}

$\bullet$ For the {\bf scalar term} $\bm{h_{00}}$ the solution of (\ref{eq8}) is the solution of the Laplace equation in the spherical coordinates, introduced in Section \ref{sec5}. The solution is
\begin{equation}
h_{00}(r,\theta) = 2\sum_{n=0}^\infty\frac{M_n}{r^{n+1}} P_n(\cos\theta)\label{eqh00}
\end{equation}
where $M_n$ are arbitrary constants, representing the static multipole moments of Thorne \cite{Thorne1,Thorne2}, and $P_n(x)$ is the Legendre polynomial of order $n$.

\noindent$\bullet$ For the {\bf vectorial $\bm{h_{0j}}$ terms} the solution of (\ref{eq9})  is
\begin{equation}
h_{0j}(r,\theta) = 2\sum_{n=1}^\infty\frac{J_n}{r^{n+1}} P_n^1(\cos\theta) e_{\varphi j}\label{eqh0i}
\end{equation}
where $J_n$ are arbitrary constants, representing the dynamic multipole moments of Thorne, $e_{\varphi}:=(-\sin\varphi,\cos\varphi,0) $ is the tangent vector associated to the $\varphi$ coordinate and $P_n^1(x)$ is the Legendre function of the first kind. 
 
\noindent$\bullet$ For the {\bf tensorial $\bm{h_{ij}}$ terms} the solution of (\ref{eq10})  is
\begin{equation}
h_{ij} = h_{00} \delta_{ij} + \partial_i w_j+\partial_j w_i\label{eqhij}
\end{equation}
where
\begin{equation}
w_j:= \sum_{n=0}^\infty\frac{Q_n}{r^{n+1}} P_n(\cos\theta) e_{z j} + \sum_{n=1}^\infty\frac{R_n}{r^{n+1}} P_n^1(\cos\theta) e_{\rho j}
\label{eq59}
\end{equation}
and $Q_n$ and $R_n$ are arbitrary constants, $e_{\rho}:=(\cos\varphi,\sin\varphi,0) $ is the tangent vector associated to the $\rho$ cylindrical coordinate and $e_{z}:=(0,0,1)$ is the tangent vector associated to the $z$ cylindrical coordinate.

It is clear that the solution of the homogeneous part of the tensor equation (\ref{eqhij}) is {\it pure gauge} \big[see (\ref{eqinf3})\big]. That is, it could be interpreted as a first order infinitesimal change of coordinates  with $\xi^i(x) := -w^i(x)$; furthermore it would be a ``harmonic change" because it is easily verified that $\Delta w^i=0$. On the other hand the complete equation, that contain the derivative of the scalar equation solution, has the simple particular solution  $h_ {00} \delta_ {ij}$. 

Now assuming that the first order solution contains this gauge dependence in the tensor term we would like to know how this gauge terms, $w_i$, propagate to the  second order term of the deviation, $q_{\alpha\beta}$. To simplify the resulting equations we are going to write down only the terms containing the first order gauge.

In particular, we would like to verify that the resulting terms coincide with those that come from a infinitesimal change of coordinates up to the second order.

\subsection{Second order in $\la$}
The second order equations to solve are (\ref{eq11}),(\ref{eq12}) and (\ref{eq13}) with ${\cal T}^{(n)}=0$ and $n=2$, then we need to use equation  (\ref{eqPN}) where only the deviation to first order in $\la$ appears. 

Let us assume the gauge dependence in the first order deviation given by equations (\ref{eqh00}), (\ref{eqh0i}) and (\ref{eqhij}), i.e. only the terms $h_{ij}$ depend on the first order gauge. From (\ref{eqmetinv}) we have $k^{\mu\nu}=-\eta^{\mu\rho}\eta^{\nu\sigma}h_{\rho\sigma}$ then in $k^{\mu\nu}$ only the terms $k^{ij}$ depend on the first order gauge. With this gauge dependence let us compute the terms $\Gamma^{(1)}$ 
\begin{equation}
\Gamma^{(1)}_{0,00} = 0,\quad \Gamma^{(1)}_{0,0j} = \frac12\partial_j h_{00},\quad
\Gamma^{(1)}_{0,ij} = \frac12\big(\partial_i h_{0j} + \partial_j h_{0i}\big)
\end{equation}
\begin{equation}
\Gamma^{(1)}_{k,00} = - \frac12\partial_k h_{00},\quad \Gamma^{(1)}_{k,0j} = \frac12\big(\partial_j h_{0k} - \partial_k h_{0j}\big)
\end{equation}
\begin{equation}
\Gamma^{(1)}_{k,ij} = \frac12\big(\delta_{kj}\partial_i h_{00} + \delta_{ki}\partial_j h_{00} - \delta_{ij}\partial_k h_{00}\big) + \partial_{ij}w_k
\end{equation}
Only the terms $\Gamma_{i,jk}$, $h_{ij}$ and $k^{ij}$ depend on the gauge, now we can write the gauge dependent terms splitting the space and time components of $P^{(2)}$ and $N^{(2)}$ in (\ref{eqPN}) and with the sign $\simeq$ we mean that we omit the terms with no gauge dependence.
\begin{equation}
P^{(2)}_0 = k^{\lambda\mu}{\Gamma}^{(1)}_{0,\lambda\mu}\simeq -  \frac12\big(\partial^k w^l+\partial^l w^k\big)\big(\partial_k h_{0l} + \partial_l h_{0k}\big)\label{eqP0}
\end{equation}
\begin{equation}
P^{(2)}_k = k^{\lambda\mu}{\Gamma}^{(1)}_{k,\lambda\mu} \simeq   -\big(\partial_iw_k+\partial_k w_i\big)\partial^i h_{00}+\partial_i w^i\partial_k h_{00}-2\partial^iw^j\partial_{ij}w_k \label{eqP2j}
\end{equation}
where we have taken into account that $\Delta w^i=0$
\begin{equation}
{N}^{(2)}_{00} = k^{\lambda\mu}\partial_\lambda{\Gamma}^{(1)}_{\mu,00} - {\Gamma}^{(1)\lambda}_{0\mu}{\Gamma}^{(1)\mu}_{0\lambda} \simeq \partial^iw^j\partial_{ij}h_{00} \label{eqN200}
\end{equation}

\begin{eqnarray}
{N}^{(2)}_{0j} =& &\; k^{\lambda\mu}\partial_\lambda{\Gamma}^{(1)}_{\mu,0j} - {\Gamma}^{(1)\lambda}_{0\mu}{\Gamma}^{(1)\mu}_{j\lambda}\simeq\nonumber
\\[1ex]
& &-\frac12\big(\partial^kw^l+\partial^l w^k\big)\big(\partial_{kj}h_{0l}-\partial_{kl}h_{0j}\big)
-\frac12\big(\partial^lh^k_0 - \partial^kh^l_0\big)\partial_{jk}w_l\label{eqN20j}
\end{eqnarray}
\begin{eqnarray}
{N}^{(2)}_{ij} = && k^{\lambda\mu}\partial_\lambda{\Gamma}^{(1)}_{\mu,ij} -{\Gamma}^{(1)\lambda}_{i\mu}{\Gamma}^{(1)\mu}_{j\lambda}-\frac12\left(k^{\lambda\mu} \partial_{ij} h_{\lambda\mu}+\partial_{i} h_{\lambda\mu}\partial_{j} k^{\lambda\mu}\right) \simeq\nonumber
\\[1ex]
&& \delta_{ij}\partial_{kl}h_{00} \partial^kw^l  + \partial_{ij}h_{00} \partial_k w^k-\partial_{(ik}h_{00}\big(\partial_{j)} w^k+\partial^k w_{j)}\big)
\nonumber
\\[1ex]
&&+\partial^k h_{00}\big(\partial_{k(i}w_{j)} - \partial_{ij}w_k\big) +\partial_{(i}h_{00}\partial_{j)k}w^k +\partial_{ik}w_l\partial^k_jw^l
\label{eqN2ij}
\end{eqnarray}
where as usual $(i\; j)$ mean simmetryzation on the index $i,j$.

Once we have determined the gauge dependent terms of the right hand side of the equations we are going to find the second order solution. As we did before let us begin  solving the scalar equation
\begin{equation}
\Delta q_{00}= 2{N}^{(2)}_{00}\simeq 2 \partial^i w^j\partial_{ij}h_{00} 
\end{equation}
The solution would be equal to the homogeneous solution, formally identical to the first order more a particular solution of the inhomogeneous equation. But, the second exact right hand side of the equation is the sum of the part that does not contain the vectors $w^k $ plus the part that contains them. Then it turns out then that the particular solution can be written as the sum of a canonical particular solution plus a pure gauge particular solution. Therefore, for our purposes, we need to find out a particular solution of the differential equation:
\begin{equation}
\Delta q_{00}=  2 \partial^i w^j\partial_{ij}h_{00}
\end{equation}
taking into account that $h_{00}$ and $w^j$ are solution of the Laplace equation then 
\begin{equation}
\Delta\big(w^i\partial_i h_{00}\big) = 2\partial^jw^i\partial_{ji} h_{00}
\end{equation}
Thus, the particular solution $w^i\partial_i h_{00}$ has exactly the structure required by the propagation of the first order gauge to second order (\ref{eqinf4}), remember that $w_j:=-\xi_j$.

Let us now solve the vector equations
\begin{eqnarray}
\Delta q_{0j}&=& 2 N^{(2)}_{0j} - \partial_j {P}^{(2)}_0  \simeq 2\partial^k w^l \partial_{kl} h_{0j} + 2\partial_{jk} w^l \partial^k h_{0l}\nonumber \\[1ex]
\partial^j q_{0j}& =&-{P}^{(2)}_0 \simeq \partial^k w^l\big( \partial_k h_{0l} +  \partial_l h_{0k}\big)
\end{eqnarray}
like in the scalar equation the homogeneous solution is formally the same as the first order one and taking into account that $\Delta w_k=0$ and $\partial^j h_{0j}=0$ we obtain that a  pure gauge particular solution is
\begin{equation}
w^k\partial_k h_{0j} + h_{0k}\partial_j w^k
\end{equation}
i.e., the solution required by the propagation of the first order gauge to second order as is given in equation (\ref{eqinf5}).

The tensor equations are
\begin{equation}
\Delta q_{ij}= 2{N}^{(2)}_{ij} - \partial_i {P}^{(2)}_j-\partial_j {P}^{(2)}_i,\quad
\partial^j q_{ij} -\frac12\partial_i \hat q=-\frac12\partial_i q_{00}-{P}^{(2)}_i
\label{tenseq}
\end{equation}
where
\begin{align}
\hspace*{-2em}&2{N}^{(2)}_{ij} - \partial_i{P}^{(2)}_j-\partial_j {P}^{(2)}_i \simeq  2\partial^kw^l \partial_{kl}h_{00} \delta_{ij} +2\partial^k h_{00}\big(\partial_{ki} w_j + \partial_{kj} w_i\big)\notag
\\[1ex]
&+ 2\partial_{ik}w_l \partial_j^k w^l+2\big(\partial^{kl} w_i\partial_{jk} w_l+ \partial^{kl} w_j\partial_{ik} w_l\big)+2\partial^k w^l\big(\partial_{kli} w_j + \partial_{klj} w_i\big)
\end{align}
and using the result of the scalar equation we have
\begin{align}
-\frac12\partial_i q_{00} - {P}^{(2)}_i  \simeq &-\frac12\partial_iw^k \partial_k h_{00} - \frac12w^k\partial_{ki} h_{00} + 2\partial^k w^l\partial_{kl}w_i
\notag
\\[1ex]
&+\partial^k h_{00}\big(\partial_k w_i + \partial_i w_k\big) - \partial_k w^k \partial_i h_{00}\label{eq75}
\end{align}
Following the same reasoning we have done above, the particular solution depending on the first order gauge term of the differential system should be the gauge term transferred from the first to the second order (\ref{eqinf5}), i.e.
\begin{equation}
q_{ij} \simeq w^k\partial_k\tilde h_{ij} + \tilde h_{kj}\partial_i w^k + \tilde h_{ik}\partial_j w^k - \partial_i w^k\partial_jw_k\label{eq76}
\end{equation}
where 
\begin{equation}
\tilde h_{ij} = h_{00}\delta_{ij} +  \partial_i w_j + \partial_j w_i\label{eq77}
\end{equation}
and it is easy to see that is a particular solution of equation (\ref{tenseq}), let us be think that this is only the particular solution depending on the first order gauge, the terms depending on the rest of the first order solution that appear in $P$ and $N$ and the terms depending on the gauge of second order are not computed.


\subsection{The possible validity of the result to the third order}
If we replace \eqref{eq77} in \eqref{eq76}, after a short calculation we obtain
\begin{align}
q_{ij}^* =\, &w^k\partial_k h_{00}\,\de_{ij}+ h_{00}\big(\partial_iw_j + \partial_jw_i\big) + \partial_iw_k \partial_jw^k\notag
\\[1ex]
 &\partial_i\big(w^k\partial_k w_j\big) + \partial_j\big(w^k\partial_k w_i\big)
 \end{align}
Therefore the exact solution of \eqref{tenseq} can be written as follows%
 \begin{align}
 q_{ij} =\,& q^c_{ij} +  w^k\partial_k h_{00}\,\de_{ij}+ h_{00}\big(\partial_iw_j + \partial_jw_i\big) + \partial_iw_k \partial_jw^k\notag
 \\[1ex]
 &+\partial_i\Big(w^{(2)}_j+w^k\partial_k w_j\Big) + \partial_j\Big(w^{(2)}_i+w^k\partial_k w_i\Big)
 \end{align}%
where $q^c_{ij}$ represent a particular solution of \eqref{tenseq} with $w^i=0$ and where the sum $\partial_iw^{(2)}_ j+\partial_jw^ {(2)}_i$ is the general solution of the homogeneous system, therefore  $\Delta w^{(2)}_i = 0 $, i.e. with $ w^{(2)}_i$ formally identical to \eqref{eq59}.
 
The second line of \eqref{eq75} could be interpreted as the main part of an infinitesimal change of coordinates to second order. It is also easy to see that $\zeta_i =\xi^{(2)}_i + w^k\partial_k\xi_i $ is the general solution of the second equation of \eqref{eq49} with $ \xi^i = -w^i $ and $\xi^{(2)}_i = w^{(2)}_i$, that is it would be an harmonic coordinates change. This suggests that the results concerning to the gauge transferred would also be valid to the third order.

\section{Interior solution in harmonic coordinates for\\ a rigidly rotating perfect fluid  with a linear\\ equation of state}
We are going to use the solution obtained in \cite{cugmmr}.

From the metric equation (\ref{eqmetrica}) and
\begin{equation}
e_{ri}\,dx^i =dr 
,\quad e_{\theta i}\,dx^i =r d\theta
,\quad e_{\varphi i}\,dx^i =r\sin\theta\, d\varphi.
\end{equation}
we have
\begin{equation}
ds^2 = \gamma_{tt}dt^2 + 2\gamma_{t\varphi} e_{\varphi j} dt dx^j
+ (\gamma_{rr}e_{ri}e_{rj}+2\gamma_{r\theta}e_{ri}e_{\theta j}+\gamma_{\theta\theta}e_{\theta i} e_{\theta j}+\gamma_{\varphi\varphi}e_{\varphi i}e_{\varphi j})dx^i dx^j
\end{equation}
that is
$$
g_{ij}:=\gamma_{rr}e_{ri}e_{rj}+2\gamma_{r\theta}e_{ri}e_{\theta j}+\gamma_{\theta\theta}e_{\theta i} e_{\theta j}+\gamma_{\varphi\varphi}e_{\varphi i}e_{\varphi j},\quad g_{0j}:=\gamma_{t\varphi} e_{\varphi j}
$$
We assume that the source of the gravitational field is a perfect fluid,
\begin{equation}
{\cal T}_{\alpha\beta} := T_{\alpha\beta} -\frac12T g_{\alpha\beta}=(\mu+p)u_\alpha u_\beta + \frac12(\mu-p)g_{\alpha\beta}
\end{equation}
whose density $\mu$  and  pressure $p$ depend only on $r$ and $\theta$ coordinates. We also assume that the fluid has no convective motion, so its velocity lies on the plane spanned by the two Killing vectors
\begin{equation}
\mbox{\boldmath$u$} = \psi(\partial_t +\omega\partial_\varphi)=\psi\big(\partial_t +\omega r\sin\theta\, e_\varphi^i\partial_i\big)
\end{equation}
from now on we are going to assume that the fluid rotates rigidly $\omega={\rm constant}$ and
\begin{equation}
\psi \equiv \left[-\left(\gamma_{tt}+2\omega\,\gamma_{t\varphi}\,r\sin\theta+\omega^2
\,\gamma_{\varphi\varphi}\,r^2\sin^2\theta\right)\right]^{-\frac12}
\label{eqnorma}
\end{equation}
is a normalization factor.
In order to obtain a postminkowskian development for the metric we need to have the density $\mu$ of order $\la$ i.e. $\mu =\la\tilde\mu$.

In the postminkowskian approximation we have
\begin{equation}
g_{00} = -1 + \la h_{00} + \la^2 q_{00},\quad g_{0j} = \omega\big(\la h_{0j} + \la^2 q_{0j}\big),\quad
g_{ij} = \delta_{ij} + \la h_{ij} + \la^2 q_{ij}
\end{equation}
this implies that
\begin{equation}
\begin{aligned}
& \gamma_{tt} = -1 + \la h_{tt} + \la^2 q_{tt},\quad
\gamma_{t\varphi} =\omega(\la h_{t\varphi} + \la^2 q_{t\varphi}),\quad \gamma_{\varphi\varphi} = 1+ \la h_{\varphi\varphi} + \la^2 q_{\varphi\varphi}\\[1.5ex]
&\gamma_{rr} = 1 + \la h_{rr} + \la^2 q_{rr},\quad
\gamma_{r\theta} =\la h_{r\theta} + \la^2 q_{r\theta},\quad
\gamma_{\theta\theta} = 1+ \la h_{\theta\theta} + \la^2 q_{\theta\theta}
\end{aligned}
\end{equation}
Let us consider Euler equations for the fluid (or, what is equivalent, the energy-momentum tensor conservation law) \cite{Boyer}
\begin{equation}
\partial_a p = (\mu + p)\partial_a\ln\psi
\qquad (a,b,\dots = r\,,\theta)\,.
\end{equation}
we are going to use the following linear equation of state (EOS)
\begin{equation}
\mu + (1-n) p =\mu_0
\end{equation}
with this EOS the Euler equations are integrated easily, to give
\begin{equation}
p = \frac{\mu}{n}\left\{\left(\frac{\psi}{\psi_\Sigma}\right)^n-1\right\},\quad \mu=\frac{\mu}{n}\left\{(n-1)\left(\frac{\psi}{\psi_\Sigma}\right)^n+1\right\}
\end{equation}
where $\psi_\Sigma$ is the value of the normalization factor $\psi$ on the surface of zero pressure,
which in turn leads to the following implicit equation for the matching surface 
\begin{equation}
p=0\quad \Longleftrightarrow \quad\psi=\psi_\Sigma\,,
\label{eqsuperficie}
\end{equation}
with this dependence on the parameter, the lowest order in $\la$ of this equation give to us cylindrical surfaces instead of spherical deformed ones unless the rotation parameter depends on  $\la$, we need $\omega^2 \propto \la$ at least then to the lowest order in $\la$ 
\begin{equation}
\psi \approx 1+\frac12\la\big(h_{tt} +\tilde\omega^2r^2\sin^2\theta\big)
\end{equation}
where we have defined $\omega^2:= \la\tilde\omega^2$ and for the constant $\psi_\Sigma$
\begin{equation}
\psi_\Sigma:=1+\la S
\end{equation}
$\om^2 = \la \tilde\om^2$ ($\tilde\om\propto\Om $, slow rotation parameter) so that this approach is consistent with the post spherical approach carried out in previous articles (see details in references \cite{cmmr} and \cite{cugmmr}). This implies that some quantities, including $g_{0j}$, have  a series expansion in half powers of $\la $. However, this situation does not affect the present work, in which we only deal with the postminkowskian approach, so we are just to consider the integer part of the orders of the series in the computation.

Then the approximate pressure is
\begin{equation}
p \approx\frac12 \la^2\tilde\mu\big(-2S+h_{tt}+\tilde\omega^2r^2\sin^2\theta\big)
\end{equation}
and for the velocity field we have
\begin{equation}
\left\{\begin{aligned}
u_0 &\approx -1+\frac12\la\big( h_{tt}-\tilde\omega^2r^2\sin^2\theta\big) 
\\[1ex]
u_j &\approx \tilde\omega\la^{1/2}\left\{r\sin\theta +\la\left[h_{t\varphi} + \frac12r\sin\theta\big(h_{tt}+2h_{\varphi\varphi}+\tilde\omega^2r^2\sin^2\theta\big)\right]\right\}e_{\varphi j}
\end{aligned}\right.
\end{equation}
now we can compute the tensor energy momentum to second order in $\la$ and we obtain
\begin{equation}
\begin{aligned}
{\cal T}_{00} &\approx \frac12\la\tilde\mu + \frac14 \la^2 \tilde\mu\big(-2(n+2)S + n h_{00}+ (n+6)\tilde\omega^2r^2\sin^2\theta\big)
\\[1ex]
{\cal T}_{0j} &\approx -\la^{3/2}\tilde\mu \tilde\omega r\sin\theta e_{\varphi j} -\frac12 \la^{5/2}\tilde\mu \tilde\omega
\\[1ex]
&\quad\Big[h_{0k}e_\varphi^k +r\sin\theta\big(-2nS +n h_{00}+2h_{kl}e_\varphi^k e_\varphi^l + (n+2)\tilde\omega^2 r^2\sin^2\!\theta\big)\Big]e_{\varphi j} 
\\[1ex]
{\cal T}_{ij} &\approx \frac12\la\tilde\mu\delta_{ij} 
\\[1ex]
&\quad+\frac14\la^2\tilde\mu\Big[ (2-n)\big(2S-h_{00}-\tilde\omega^2r^2\sin^2\!\theta\big)\delta_{ij} + 2h_{ij} + 4\tilde\omega^2 r^2\sin^2\theta e_{\varphi i}e_{\varphi j}\Big]
\end{aligned}\label{tie2order}
\end{equation}
\subsection{First order in $\la$}
The Einstein's equations to first order  are (\ref{eq8}), (\ref{eq9}) and (\ref{eq10}), the scalar equation
\begin{equation}
\Delta h_{00}=-2{\cal T}^{(1)}_{00}=-\tilde\mu
\end{equation}
has as solution, well behaved  in $r=0$ given by
\begin{equation}
h_{00}(r,\theta) = -\frac16\tilde\mu\,r^2+2\sum_{n=0}^\infty m_n r^n P_n(\cos\theta)\label{eqintesc}
\end{equation}
The vector equations are
\begin{equation}
\begin{aligned}
&\Delta h_{0k}=-2{\cal T}^{(1)}_{0k} = +2\tilde\mu\,\omega\, r\sin\theta\,e_{\varphi k}
\\[1ex]
&\partial^k h_{0k} =0
\end{aligned}\label{eqintvec}
\end{equation}
and their solution is
\begin{equation}
h_{0k}(r,\theta) =\left[\frac15\tilde\mu\,\omega\, r^3\sin\theta + 2\sum_{n=1}^\infty j_n r^n P_n^1(\cos\theta)\right] e_{\varphi k}
\end{equation}
as for the exterior solution there are not gauge dependent terms in the first order solution for $h_{00}$ and $h_{0j}$.

The tensor equations are
\begin{equation}
\begin{aligned}
&\Delta h_{ij} = -2{\cal T}^{(1)}_{ij} = -\tilde\mu\, \delta_{ij}
\\[1ex]
&\partial^k h_{kj} -\frac12\partial_j\hat h=-\frac12\partial_j h_{00}
\end{aligned}
\end{equation}
and their solution is
\begin{equation}
h_{ij} = h_{00}\, \delta_{ij} + \partial_i \bar w_j+\partial_j \bar w_i
\end{equation}
where now
\begin{equation}
\bar w_j:= \sum_{n=1}^\infty\bar Q_n r^n P_n(\cos\theta)\, e_{z j} + \sum_{n=1}^\infty\bar R_n r^n P_n^1(\cos\theta)\,e_{\rho j}
\label{galgaint}
\end{equation}
As in the exterior problem the solution of the homogeneous part of this tensor equation is {\it pure gauge}, see (\ref{eqinf3}) with $\xi_j:=-\bar w_j\,$. And the complete equation that contain the derivative of the scalar equation solution has the simple particular solution  $ h_ {00} \delta_ {ij} $ too; then we can use the same expressions (\ref{eqP0}), (\ref{eqP2j}), (\ref{eqN200}), (\ref{eqN20j}) and (\ref{eqN2ij}) we have used above for the exterior solution.

Now also we want to determine how the terms {\it ``pure gauge"\/} of the first order solution are transmitted  to the second order solution  and in particular verify that the resulting terms coincide with  those coming from an infinitesimal second order change of coordinates.


\subsection{Second order in $\la$}
We can now go to second order i.e. to solve (\ref{eq11}), (\ref{eq12}) and (\ref{eq13}) with ${\cal T}$ given by equations (\ref{tie2order}) when $n=2$. Do not confuse this $n$ related with the order of the approximation we perform with the $n$ appearing in the EOS. As in the exterior problem we are going only to look for the terms in the solution corresponding to the propagation of the first order gauge.

The scalar equation is
\begin{equation}
 \Delta q_{00}= 2{N}^{(2)}_{00}- 2{\cal T}^{(2)}_{00} \simeq 2 \partial^i \bar w^j\partial_{ij}h_{00} \quad
\end{equation}
the term ${\cal T}^{(2)}_{00}$ do not contain the gauge vectors $\bar w^i$ the terms to solve have the same aspect that in the exterior case. 
And we have
\begin{equation}
\Delta\big(\bar w^i\partial_i h_{00}\big) = \Delta \bar w^i\,\partial_i h_{00} +2\partial^j\bar w^i\,\partial_j(\partial_i h_{00}) + \bar w^i\Delta(\partial_i h_{00})= 2\partial^j\bar w^i\,\partial_{ji} h_{00}
\end{equation}
where we have taking into account that
\begin{equation}
\Delta \bar w^i = 0\ ,\quad  \Delta(\partial_i h_{00})=\partial_i\big(\Delta h_{00}\big) =0
\end{equation}
Then, the particular solution for the terms depending on the first order gauge has the aspect required for a gauge change to second order (\ref{eq7}).

For the vector equations the term ${\cal T}^{(2)}_{0j}$ only is gauge depending in the $h_{ij}$ term, that is
\begin{equation}
 {\cal T}^{(2)}_{0j}\simeq -2\tilde\mu\omega r\sin\theta\big(\partial_k\bar w_l e_\varphi^k e_\varphi^l\big)e_{\varphi j}
\end{equation}
the equations to find a particular solution of the gauge depending part are
\begin{equation}
\begin{aligned}
\Delta q_{0j}&= 2 {N}^{(2)}_{0j} - \partial_j{P}^{(2)}_0 - 2{\cal T}^{(2)}_{0j}
\\[1ex]
& \simeq 2\partial^k\bar w^l \partial_{kl} h_{0j} + 2\partial_{jk}\bar w^l \partial^k h_{0l}+4\tilde\mu\omega r\sin\theta\big(\partial_k\bar w_l e_\varphi^k e_\varphi^l\big)e_{\varphi j}
\\[1ex]
&\partial^j q_{0j} =-{P}^{(2)}_0\,\, \simeq \partial^k\bar w^l\big( \partial_k h_{0l} +  \partial_l h_{0k}\big)
\end{aligned}
\end{equation}
We know from  the second order gauge infinitesimal transformation (\ref{eq7}) that the expression $\bar w^k\partial_k h_{0j} + h_{0k}\partial_j \bar w^k$ should be the right answer,  let's try this particular solution  
\begin{align}
\Delta\big(w^k\partial_k h_{0j} + h_{0k}\partial_j w^k\big)=& 2\partial^l\bar w^k\,\partial_{lk}h_{0j} + 2\partial^l h_{0k}\,\partial_{lj}\bar w^k\notag
\\[1ex]+\bar w^k \partial_k \Delta h_{0j}+ \Delta h_{0k} \partial_j \bar w^k
&  
\end{align}
and using the first order vector equation (\ref{eqintvec}) and the expression for $\bar \omega_k$ (\ref{galgaint}) we get
$$ \bar w^k \partial_k \Delta h_{0j}+ \Delta h_{0k} \partial_j \bar w^k=4\tilde\mu\omega\sum_{n=1}^\infty \bar R_n r^n P_n^1(\cos\theta)\,e_{\varphi j}
$$
and the last term in the expression for $\Delta q_{0j}$ using (\ref{galgaint}) 
\begin{equation}
r\sin\theta\big(\partial_k\bar w_l e_\varphi^k e_\varphi^l\big)=\sum_{n=1}^\infty \bar R_n r^n P_n^1(\cos\theta)
\end{equation}
Now if we put this particular solution in the harmonicity condition
\begin{equation}
\partial^j\big(\bar w^k\partial_k h_{0j} + h_{0k}\partial_j \bar w^k\big) = \partial^j\bar w^k\big( \partial_j h_{0k} +  \partial_k h_{0j}\big)
\end{equation}
then the gauge propagation gives the particular solution to second order for the vector equations too.

To finish with let us solve the following tensor equations
\begin{equation}
\Delta q_{ij}= 2{N}^{(2)}_{ij} - \partial_i {P}^{(2)}_j- \partial_j {P}^{(2)}_i - 2  {\cal T}^{(2)}_{ij},\quad
\partial^j q_{ij} -\frac12\partial_i \hat q=-\frac12\partial_i q_{00}-{P}^{(2)}_i
\end{equation}
the only difference with the equations we have solved for the exterior problem is the term
\begin{equation}
 {\cal T}^{(2)}_{ij}\simeq \frac12\tilde \mu\big(\partial_i\bar w_j + \partial_j\bar w_i\big)
 \end{equation}
and the remaining terms are determined from (\ref{eqN2ij}) and (\ref{eqP2j}) and the solution of the scalar equation (\ref{eqintesc}).

Then as we did above let's try with the particular solution
\begin{equation}
q_{ij}^* = \bar w^k\partial_k h_{ij} +  h_{kj}\partial_i \bar w^k +  h_{ik}\partial_j \bar w^k - \partial_i \bar w^k\,\partial_j\bar w_k
\end{equation}
where
\begin{equation}
h_{ij} = h_{00}\delta_{ij} +  \partial_i \bar w_j + \partial_j \bar w_i
\end{equation}
As in the exterior vacuum  problem a direct calculation shows that this is indeed solution.

\section{Conclusions}
When we solve the Einstein equations in harmonic coordinates to each order we obtain solutions that depend on the  development parameter and some constants, we can distinguish to types of constants multipole moments and the so called gauge constants, giving all solutions but keeping them within the family of harmonic coordinates. These gauge constants are essential if we match the interior solution to the exterior solution to get a global solution but with this procedure we obtain messy solutions when we compute the nonlinear terms in the development constant $\lambda$.

It seems natural to think that an infinitesimal change of harmonic coordinates depending on the same constant $\lambda$ allows us to transform the metric and simplify the solutions so that we can manipulate solutions that have not gauge constants and obtain the general solution by performing an infinitesimal coordinate transformation.

This paper demonstrates that this happens to at least second order in the constant $\lambda$.

It  is to say, we can find a first order solution with all the gauge constants then  with this solution we can generate the second order solution which will contain quadratic terms in the first order gauge constants and on the multipolar first order constants and linear terms on the second order gauge constants, in \cite{cmmr}, \cite{cugmmr} we do this following the procedure sketched in the appendix, let's call this solution $\bar g$ or we can find the first order solution with the first order gauge constants equal to zero and then with this solution generate the second order solution that only contain multipole terms.  Then we can transform this solution without gauge terms by a  second order infinitesimal coordinate transformation, with the condition that transforms from one set of harmonic coordinates to another set of harmonic coordinates,   \eqref{eq40}, let's call this solution $\hat g$ then with a redefinition of the gauge constants and of the second order multipole moments we can made to coincide the solutions $\bar g$ and $\hat g$.\footnote{The solution $\bar g$ depend on the procedure of how we construct it because a homogeneous solution can be hidden in it, if two solutions obtained by two different procedures do not coincide the difference between them is a solution of the homogeneous differential system of equations}

This last procedure, without gauge constants, to find the solution let us to get simpler solutions and saves computing time, then if we we need to match them with another interior or exterior solution we must perform a second order infinitesimal harmonic coordinates transformation on the solution without gauge constants to obtain a solution with all the constants needed to do the Lichnerowicz matching of solutions.

\end{document}